\documentclass[12pt]{spieman}  % 12pt font required by SPIE;
\usepackage{amsmath,amsfonts,amssymb}
\usepackage{graphicx}
\usepackage{setspace}
\usepackage{tocloft}
\usepackage{lineno}
\usepackage{xspace} % for xspace

\newcommand{\sh}{SLEDGEHAMMER\xspace}

\title{Design of a Mission to Measure the Shape and Substructure of the 511 keV $\gamma$-Ray Line from the Center of the Milky Way}

\author[a,$\dagger$]{Kun Hu}
\author[a]{Matthew Fritts}
\author[b]{Daniel Becker}
\author[c]{Daniel Schmidt}
\author[a]{Fang Zhou}
\author[d]{Adrika Dasgupta}
\author[d]{Fabian Kislat}
\author[c]{Mark Keller}
\author[a]{Sohee Chun}
\author[a,e]{Argen Gian Detoito}
\author[a]{Ephraim Gau}
\author[f,c]{Xiaoyue Jin}
\author[c]{Douglas Bennett}
\author[a]{Dana Braun}
\author[b]{Johnathon Gard}
\author[c]{John A. B. Mates}
\author[b]{Jason Nobles}
\author[a]{Danny Radomski}
\author[a]{Nicole Rodriguez Cavero}
\author[c]{Ryan Snodgrass}
\author[c]{Daniel Swetz}
\author[c]{Joel Ullom}
\author[a]{Shravan Vengalil Menon}
\author[b]{Joel Weber}
\author[d]{Kasun Wimalasena}
\author[a,e,$\ddagger$]{Henric Krawczynski}
\affil[a]{Washington University in St. Louis, Department of Physics and McDonnell Center for the Space Sciences, St. Louis, MO 63130, United States}
\affil[b]{University of Colorado, Boulder, CO 80309, United States}
\affil[c]{National Institute of Standards and Technology, Boulder, CO 80305, United States}
\affil[d]{University of New Hampshire, Space Science Center, Department of Physics and Astronomy,
Durham, NH 03824, United States}
\affil[e]{Washington University in St. Louis, Center for Quantum Leaps, St. Louis, MO 63130, United States}
\affil[f]{Theiss Research, La Jolla, CA 92037, United States}

\cftpagenumbersoff{figure}
\cftpagenumbersoff{table}

\begin{document} 
\maketitle

\begin{abstract}
The 511 keV electron-positron annihilation feature near the galactic center has been detected for more than half a century, yet its origin remains a mystery.
In this paper, we describe a concept for a balloon-borne 511 keV 
$\gamma$-ray mission called the 
511-Spectrometer Mission. 
The mission will use Transition-Edge Sensor (TES) arrays with thick metal absorbers that are thermally coupled to the TES. The strength of the approach is a projected 
energy resolution of 200 eV Full Width Half Maximum (FWHM) at 511\,keV, enabling detailed studies of the shape and substructure of the 511 keV emission from the galactic center region. 
A first mission equipped with 8,192 $\gamma$-ray detectors and a fully active shield and collimator could detect the galactic center with ~35 $\sigma$ statistical significance. 
We present the mission concept as well as first results
obtained with a prototype detector equipped with $1.35\times1.35\times2$\,mm$^{3}$ Bi absorbers.
The detector has a quantum efficiency of 15\% for 511\,keV photons in photoelectric effect interactions.
In tests with a $^{137}$Cs source, these prototype detectors show an energy resolution of 525\,eV FWHM at 662\,keV.  
We end with a discussion of follow-up missions that use coded mask imaging, or use concentrating or focusing optics to 
scrutinize the sources of 511\,keV $\gamma$-rays on 
smaller angular scales. 
\end{abstract}

% Include a list of up to six keywords after the abstract

% Include email contact information for corresponding author
{\noindent \footnotesize\textbf{$^\dagger$ }Kun Hu,  \linkable{hkun@wustl.edu} \quad \textbf{$^\ddagger$}Henric Krawczynski,  \linkable{krawcz@wustl.edu}}

\begin{spacing}{1}   % use double spacing for rest of manuscript

\section{Introduction}
\label{sect:intro}
Johnson, Harnden \& Haymes discovered the first evidence for the 511\,keV $\gamma$-ray emission from the galactic center region with an experiment flown in 1970 on a stratospheric balloon flight \cite{1972ApJ...172L...1J}.
In the meantime, the signal has been scrutinized by several generations of $\gamma$-ray experiments, with the most constraining spectral and imaging data coming from the {\it CGRO/OSSE}, {\it INTEGRAL/SPI} and the 
{\it Compton Spectrometer and Imager (COSI)} missions    \cite{1997ApJ...491..725P,2005A&A...441..513K,2005MNRAS.357.1377C,2008Natur.451..159W,2010ApJ...720.1772B,2011MNRAS.411.1727C,2020ApJ...895...44K,2020ApJ...897...45S}. 
The spectral analysis of the {\it INTEGRAL} and {\it COSI} data suggests a large positronium fraction due to the existence of a continuous spectral component below 511 keV (generated by the $3\gamma$ decay of ortho-positronium). 
%This indicates
\textcolor{black}{The high positronium fraction (with binding energy $\sim$ 6.8 eV) and the narrow 511 keV line width together indicate} 
that the kinetic energies of the positrons must be smaller than a few eV before annihilation. 
Furthermore, the large bulge-to-disk flux ratio from the imaging analysis constrained the spatial distribution of the positron sources. 
Many different scenarios have been invoked to explain the 511 keV emission. The $\gamma$-rays may be produced by the annihilation of 
positrons from the decay of dark matter particles 
with electrons from the interstellar medium \cite{2018MNRAS.479.2229C}. 
Alternatively, the electrons and positrons might be created 
in X-ray binaries or other astrophysical sources, 
involving exclusively standard model physics \cite{2011RvMP...83.1001P}. 

Dark matter can create positrons through decay \cite{2004PhRvD..70f3506H} and annihilation \cite{2004PhRvL..92j1301B} of $\sim$MeV particles, or 
de-excitation of dark matter in models with hidden sectors \cite{2007PhRvD..76h3519F}. 
The various models predict distinct spatial distributions for the resulting positrons. Dark matter decay leads to a local positron creation rate $\dot{n}_{\rm e+}$ proportional to the number density of the dark matter particle $n_{\rm DM}$ and thus to a 511 keV flux map mimicking the dark matter density profile in the galaxy. Conversely, dark matter annihilation or de-excitation leads to $\dot{n}_{\rm e+}\propto \langle\sigma v\rangle n^2_{\rm DM}$, where $\langle\sigma v\rangle$ is the 
velocity-weighted cross section for the annihilation or excitation. This produces a somewhat steeper positron distribution due to the presence of $n^2_{\rm DM}$. Then the expected 511 keV flux map can be obtained through integration of the positron production rate over the line of sight with a prescribed density profile and velocity distribution.

{Astrophysical (rather than particle physics) explanations} of the 511\,keV emission focus on the production of positrons in astrophysical sources. Low-energy positrons can be generated via $\beta^+$ decay of $^{26}$Al, $^{44}$Ti or $^{56}$Ni, which are created in massive stars, core-collapse supernovae or Type Ia supernovae. The observation of the 1.8 MeV $^{26}$Al decay line confirmed the existence of this radioactive decay channel in the galaxy and also suggested a large abundance of radioactive isotopes in the galactic disk \cite{1982ApJ...262..742M,1995A&A...298..445D}. Therefore it is believed that $\beta^+$ decay of $^{26}$Al should contribute at least a certain portion of the positrons for the 511 keV emission. The distributions of massive stars or supernovae can hardly explain the emission morphology and the bulge-to-disk flux ratio. Large amounts of $e^{+/-}$ pairs can also be generated in the disk, corona, or the jets of X-ray binaries via $\gamma-\gamma$ pair creation, or through 
pair cascades
in the magnetospheres of neutron stars \cite{2015ApJ...807..130V}.
The created $e^{+/-}$ can be injected into the interstellar medium through X-ray binary jets or the open field lines of pulsars and contribute to the galactic 511 keV emission.
Some of the $e^{+/-}$ pairs are expected to annihilate near the system where they are created, making them potential point-like sources of 511 keV emission \cite{2006A&A...457..753G}. %\\[0.4ex]
% {\bf Positron propagation:} 

Regardless of the origin of the positrons, the propagation of the $e^+$ in the interstellar medium (ISM) adds complexity to the picture.  It might be dominated by collisional interactions between (charged or neutral) particles \cite{2014A&A...564A.108A}, or the  deceleration via scattering off the magnetohydrodynamic waves in the interstellar medium \cite{2009ApJ...698..350H}. 
The positrons may 
not propagate far away from their sources \cite{2014A&A...564A.108A}.
The existing data suggest that the annihilating $e^+-e^-$ pairs
have eV-energies (see \cite{2020ApJ...895...44K} and references therein). If so, bright regions could emit narrow 511\,keV lines with frequency shifts given by the bulk motion of the plasma and with widths that are much smaller than the energy resolution of the 511-Spectrometer mission.
% \begin{redbox}

In this paper we describe a balloon-borne mission, the 511-Spectrometer Mission (511SM), that uses Transition-Edge Sensors (TES) microcalorimeter arrays. These detectors promise a 511\,keV 
energy resolution between $\sim$200\,eV (goal) and $\sim$525\,eV (currently demonstrated in the lab) FWHM. 
Such a mission would be complimentary to the 
{\it COSI} Small Explorer ({\it COSI-SMEX}) satellite mission, currently being developed for a possible launch in 2027. {\it COSI-SMEX} uses an array of High Purity Germanium (HPGe) detectors with an energy resolution requirement of 6\,keV FWHM at 511\,keV. 
{\it COSI-SMEX} can image the 511\,keV line across the Milky Way, and will obtain energy spectra for different emission regions \cite{2023arXiv230812362T}. 

The 511SM will have 10 to 30 times better energy resolution than {\it COSI}, enabling precision measurements 
of the line shape 
and the substructure of the 511\,keV $\gamma$-ray line. The {\it INTEGRAL} and {\it COSI} experiments measured intrinsic 511\,keV line widths of 5.7\,keV\,$\pm$\,0.33\,keV and 4.00\,keV\,$\pm$\,0.94 keV, respectively.  
The detailed analysis of the {\it INTEGRAL} line indicates the presence of a narrow 1.41\,keV FWHM and a broad 5.4\,keV FWHM component. Theoretically, the width and shape of the 511\,keV line depend on the annihilation environment with different results depending on the density and type of the target material with which the positrons interact. Energy resolutions surpassing those achievable with HPGe detectors are required to measure the line shape.
All else being equal, the %precision 
\textcolor{black}{instrumental resolution} 
with which a mission can measure the line-of-sight velocity of emitting plasma is given by %$\Delta E c  / (E\,\sqrt{n})$
\textcolor{black}{$\Delta E c  / (E)$}, where $\Delta E$ is the energy resolution at photon energy $E$, $c$ is the speed of light\textcolor{black}{.} %, and $n$ is the number of detected photons. 
Thus, a mission with $\Delta E = 200$\,eV FWHM would measure line-of-sight velocities (LoS velocities) by a factor of $\sim$30 better than a mission with $\Delta E =$ 6\,keV FWHM. Crucially, the 511SM LoS velocity resolution of between 120 and 330 km/s FWHM is smaller than the stellar velocity dispersion in the Milky Way \cite{2018A&A...616A..83V}, enabling the association of a point source with specific Milky Way components, such as spiral arms or the bulge, based on LoS velocity measurements alone.
\textcolor{black}{For an unresolved line detected with $n$ photons, the statistical centroid precision improves as $1/\sqrt{n}$.}

We discuss the design of the 511SM in Section~\ref{s:design}, and the expected outcome of the observations in Sect.\,\ref{s:outcome}. 
We end with a summary and discussion of the 511SM and related missions, i.e.\,a similar mission with coded mask imaging, a pointed 511\,keV $\gamma$-ray telescope, and a pointed mission making use of $\gamma$-ray optics.     

\section{511-Spectrometer: A balloon-borne 511 keV Microcalorimeter Mission}
\label{s:design}
\subsection{Overall Design}

\begin{figure}
  \centering
\includegraphics[width=0.6\textwidth]{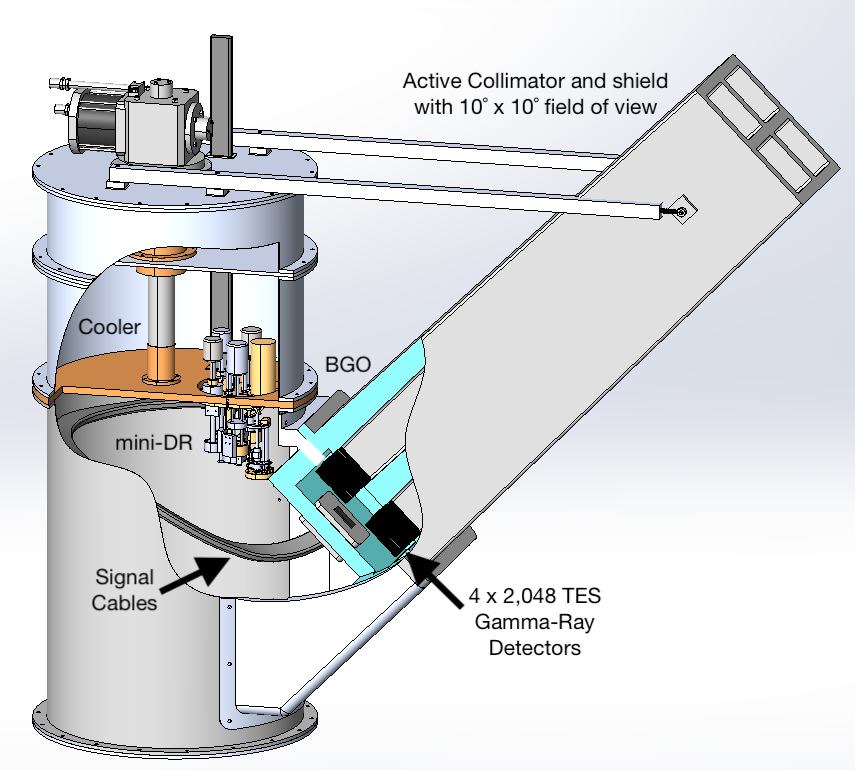}
%  [trim=-0.1cm 0 0 -0.1cm]
  \caption{Sketch of the 511-Spectrometer mission's $\gamma$-ray telescope that uses 8,192 microcalorimeter detectors with {a collimator and an active shield} to measure the width and sub-structure of the 511\,keV emission from the galactic center with an energy resolution of $\sim$ 200 eV FWHM.
  The 4.2\,K platform needed by the mini-DR could be provided by a closed-cycle cryo-cooler. \textcolor{black}{A vibration isolation system, which is not depicted in the sketch, will be implemented for the cryo-cooler.}    
  \label{f:design2}}%\vspace*{-3ex}
\end{figure}
We introduce the concept of the 511SM, a balloon-borne mission that uses 8,192 $\gamma$-ray microcalorimeter detectors to observe the galactic center with high statistical significance and unprecedented energy resolution.
Each microcalorimeter detector will use a high-$Z$ (roughly 4%$2\times1\times2$
\,mm$^3$) Bi or BiSn absorber 
which record 511\,keV $\gamma$-rays with a photoelectric effect detection efficiency of 15\%.
Scaling the energy resolution of the SLEDGEHAMMER \cite{2015ITAS...2581878B} Sn detectors (55\,eV at 98\,keV \cite{mates_etal_2017}) with the square-root of the heat capacitance of the Bi and Sn absorbers, the predicted energy resolution of the 511SM is $\sim$200\,eV FWHM. \textcolor{black}{This value represents an idealized, noise-limited projection and may be degraded by large-signal nonlinearity, non-ideal absorber thermalization, and imperfect absorber–TES thermal coupling.}

The 511-Spectrometer $\gamma$-ray telescope (Fig.\,\ref{f:design2}) consists of the cryostat-detector assembly and a 1\,m long collimator and \textcolor{black}{active} shield made of Al-encased Bismuth Germanate Oxide (BGO). The tapered collimator
with front and rear BGO thicknesses of 2\,cm and 5\,cm, respectively, limits the field of view of the detector assembly to a $\pm$5$^{\circ}$ region in the sky.
The BGO shield is outside of the cryostat where possible, but also partially inside, since surrounding the cryostat would require too much BGO. The detector inside the cryostat views the sky through a window assembly consisting of a thin plastic scintillator (used to tag charged particles), Be windows, and infrared-optical light blocking filters.

The telescope will follow the galactic center, requiring an azimuthal tracking by $\sim$1$^{\circ}$ per day, and elevation tracking. 
We can use the WASP Celestial Attitude Reference and Determination System to provide absolute pointing information \cite{stuchlik}.
The 511SM could be flown on a 30 to 100 day super-pressure balloon flight at 33.5 km (110,000 feet) altitude launched from W\={a}naka, New Zealand, or on a 8 to 30 day Long Duration Balloon flight at 38.1 km (125,000 feet) altitude launched from McMurdo, Antarctica. Based on detailed simulations of the source visibilities and atmospheric transmissivities, the sensitivities averaged over 24 hour windows turn out to be similar for both types of flights, making the longer super-pressure balloon flight the better choice. For the latter, the galactic center is $\sim$8 hours a day below the horizon. These windows can be used for the observation of other sources.

The overall design of the 511SM critically depends on the choice of the cooling system. Conventionally, liquid helium (LHe) is used to provide the 4.2\,K platform temperature \cite{gudmundsson_etal_2015,2014JAI.....340001G}.  
The LHe approach faces the problem of high LHe costs. Even a smart cryostat design would require $\sim$1000\,L of LHe for a 100-day flight \cite{2020SPIE11445E..7AL}. We would likely use a Continuous Miniature Dilution (CMD) refrigerator similar to the one used on DR-TES (Dilution Refrigerator\cite{2019MS&E..502a2134C}-Transition Edge Sensor) with $\sim$1.5\,$\mu$W cooling power at 75\,mK to cool the detector. An additional pre-cooler may be used to cool the cables and mechanical support structures to 300\,mK. 
\textcolor{black}{Vibrations from the cryocooler can raise the detector temperature and introduce electric noise. The Hitomi and XRISM missions employ vibration isolation system to mitigate the  degradation of the spectroscopic performance\cite{2018JATIS...4a1216T,2025JATIS..11d2004Y}. We have not yet decided how to solve this problem. The solution will likely involve mechanically separating the vibrating components from the cryostat and loading them with inert mass to move resonances to lower frequencies. The solution will require additional engineering and test measurements.}

A microwave multiplexing method based on Superconducting Quantum Interference Devices (SQUIDs) will be used to read out the detectors.
Assuming a multiplexing ratio of 1:512, we need a total of 16 incoming and outgoing microwave cables (SC-086/50 NbTi cables\footnote{Certain commercial products are identified to specify the experimental study adequately. This does not imply endorsement by NIST or that the products are the best available for the purpose.} from Coax Co. LTD.  \cite{coax}) dissipating a total of 0.05 nW between 75\,mK and 300\,mK and 0.05 $\mu$W between 300\,mK and 4\,K.
The detector assembly will be supported by the signal cables and low-thermal-conductivity carbon fiber arrow shafts  \cite{2018RScI...89l6105P}.
The detector modules reside inside a superconducting niobium shield and a %0.062'' 
1.6 mm thick Amumetal 4K shield.
The gradiometric design of the NIST SQUIDs\cite{2012RScI...83i3113B,noroozian_etal_2013,mates_etal_2017} reduces the effective pick up area to between 0.5\% and 2\% of their footprint. All components not touching the detectors will be held at 4.2\,K, and their radiation will be copper-shielded from the detector assembly. 

\textcolor{black}{The warm readout electronics follow a standard µmux RF chain: the resonator comb transmitted from the focal plane is amplified by a cryogenic HEMT at $\sim4$ K, followed by additional gain at an intermediate stage (e.g., $\sim50$ K LNA) and a room-temperature RF amplifier before digitization. All these amplifiers will be thermally anchored and properly shielded to prevent heating the stages below 4 K. 
For the room-temperature backend we baseline an RFSoC platform (e.g., Xilinx ZCU111, as demonstrated in the DR-TES flight), which integrates multi-GS/s ADCs/DACs with FPGA logic and embedded processors, enabling direct digitization and manipulation of GHz microwave signals.}

The 511SM achieves a remarkable signal-to-noise ratio due to the combined influence of several factors: (i) The high atomic number of Bi ($Z$=83) leads to the dominance of the photoelectric effect cross-section over the Compton cross-section up to 565 keV. As the fraction of photoelectric effect events is large (55\% at 511 keV) and 
one-pixel events tend to be signal and not background events, we rely exclusively on one-pixel events for the 511SM. For a 2\,mm thick Bi absorber, the probability of absorbing 511\,keV $\gamma$-rays in photoelectric effect interactions is 15\%. (ii) The {\it XL-Calibur} in-flight data show that the use of an active %rather than passive 
{shield} reduces the 511\,keV background by more than one order of magnitude compared to a passive shield assembly \cite{2023NIMPA104867975I,2025MNRAS.540L..34A}.  (iii) The {\it XL-Calibur} measurements show that background events tend to trigger two or more pixels rather than a single one, and a one-pixel event selection cut suppresses the background by one order of magnitude
\cite{2023NIMPA104867975I}.

\subsection{SLEDGEHAMMER $\gamma$-ray detectors}
\label{s:rd}

\begin{figure}
  \centering
\includegraphics[width=0.85\linewidth]{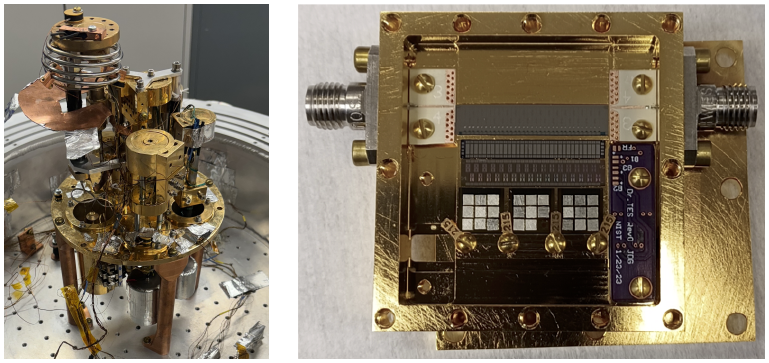}
  \caption{The DR-TES project flight-tested several technologies relevant for the 511-Spectrometer Mission on a conventional 1-day balloon flight from Fort Sumner (NM) in September, 2024. Left: Chase Research Cryogenics CMD refrigerator mounted in the flight cryostat. Right: the flight detector featuring 26 (0.3$\times$1.35$\times$1.35\,mm$^3$) Sn absorbers. 
  \label{g0}
  }%\vspace*{-3ex}
\end{figure}

Microcalorimeter detectors are already transforming the field of soft and intermediate energy X-ray spectroscopic imaging. The recently launched X-Ray Imaging and Spectroscopy Mission ({\it XRISM}) uses a microcalorimeter array with 35 pixels for the spectroscopic observations of X-ray sources over the 0.3 keV to 13 keV energy range with $<$7\,eV FWHM energy resolutions \cite{2022SPIE12181E..1SI}.  
The Athena mission \cite{2022cosp...44.2316B}, the Line Emission Mapper (LEM) \cite{2022arXiv221109827K}, and the Lynx mission \cite{2019SPIE11118E..0KS,2022JLTP..209..337D}  will use microcalorimeter arrays with $\sim$eV energy resolutions and with
1504, 13,924, and $>$100,000 TES pixel, respectively.
Microcalorimeter detectors operate typically at $\sim$100\,mK temperatures taking advantage of the small heat capacitances of the absorber materials and low phonon noise at these temperatures. 
Individual photons are recorded via the temperature transients they cause in the absorber. For each absorber, a TES voltage-biased within its superconducting-to-normal transition is used to convert the temperature transients into a current transient. Owing to the sharpness of the transition, a tiny temperature change translates into a large change of the TES current. %(Fig.\,\ref{f1}).

\begin{figure}[tb]
  \centering
  \includegraphics[width=\textwidth]{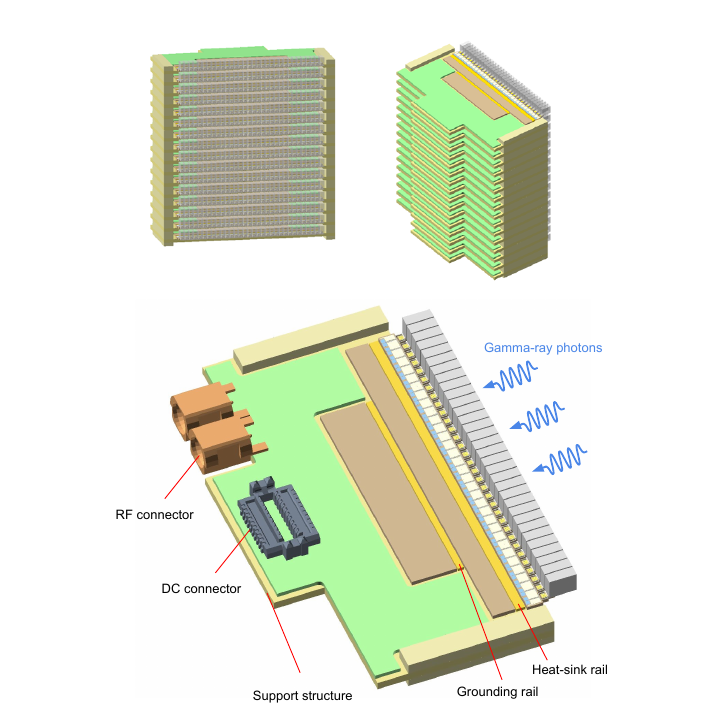}
  \caption{Top: The 511SM design uses 2048 detector modules. The images show front and rear views of an example 16$\times$32 detector assembly made of 16 1-D sub-modules (green, yellow and brown) each with 32 detectors (grey). The 511SM would use larger modules with 32$\times$64 detectors. Bottom: %\sout{Cross section of one of the 1-D detector sub-modules from Fig.\,\ref{f:module}} 
  One of the 1-D detector sub-modules showing the mounting of the absorbers (grey) and various electronic components (green and brown) on the main support structure.
  }
  \label{f:module}
\end{figure}
%%%%%%%%%%%%
\begin{figure}[tb]
  \centering
\vspace*{1cm}
\includegraphics[width=\textwidth]{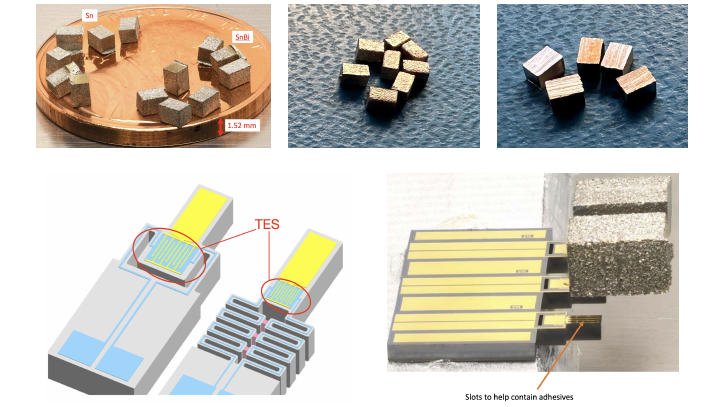}
  \caption{
Top: We fabricated 1.3$\times$1.3$\times$2 mm$^3$ absorbers using  Electrical Discharge Machining %(upper left panel: Sn, Bi, and BiSn), 
(upper left panel: Sn and BiSn), 
casting (upper center panel: Bi), and Wafer Dicing (upper right panel: Bi).  Wafer dicing gives the most uniform absorbers with the smoothest surfaces. Lower-left: The 511SM design uses large absorbers mounted on thick Si prongs. The CAD drawings show exemplary designs with straight and serpentine Si legs supporting the TES and absorber, leading to widely varying thermal couplings (Si structure: grey, electrical contacts:  blue, TESs: in red circles, and mating surface for the absorber: yellow).  Lower-right: Test carrier structures with the casted  1.35$\times$1.35$\times$2\,mm$^3$\,Bi absorbers.\vspace*{-0.05cm} }
  \label{f:unit}
\end{figure}

Our work will build on the success of the SLEDGEHAMMER $\gamma$-ray micocalorimeter array developed by NIST and the University of Colorado, Boulder \cite{2012RScI...83i3113B,mates_etal_2017}.
That array features 250 (1.45$\times$1.45 $\times$0.3\,mm$^3$) Sn absorbers thermally connected to a MoCu TES with a transition temperature $T_{\rm c}$ of  $\sim$100\,mK.
The \sh detectors achieve an energy resolution of 55\,eV FWHM at 98 keV {(Figure 5 of Mates et al. 2017~\cite{mates_etal_2017})}, and can be used from 20\,keV to 240\,keV. %\sout{, although the detection efficiency drops above 60 keV}. 
The TESs are deposited on a silicon nitride membrane that provides mechanical support and weakly couples the sensors to the bath temperature. %(Fig.\,\ref{f3}, left panel)
The Sn absorbers are glued to SU8 epoxy pillars that mechanically support the absorbers and establish the thermal link between the absorbers and the TESs.  % (Fig.\,\ref{f3}, center panel).
The TESs are read out with microwave-multiplexed SQUIDs.

We tested prototype \textcolor{black}{SLEDGEHAMMER} TES $\gamma$-ray detectors and the mini dilution refrigerator with the {DR-TES} payload. 
The {DR-TES} project consists of the Chase Research Cryogenics mini-dilution refrigerator (Fig.\,\ref{g0}, left panel) and a NIST $\gamma$-ray TES microcalorimeter (Fig.\,\ref{g0}, right panel), both mounted inside a cryostat precooled by liquid helium. The {DR-TES} completed a successful 1-day balloon flight launched from Fort Sumner, New Mexico, on September 24th, 2024. During the flight, the dilution refrigerator cooled the detector to $\sim$ 90 mK, and hard X-ray photons from an on-flight radiation source were recorded as pulses. The behaviors of the cooling system and the $\gamma$-ray detector during the flight will be detailed in different papers.

\subsection{A focal plane with 8,192 $\gamma$-ray detectors}
% \label{s:ph}

We envision the assembly of large detector arrays with the PRONGHORN approach developed by NIST and CU. This design (Fig.\,\ref{f:module}) structures the focal plane into modules, sub-modules, and TES-absorber units. 
Each module is made of identical $m$ sub-modules each holding $n$ TES-absorber units. The innovative design involves building sub-modules in which the majority of readout and heat sinking components can be placed behind the highly sensitive TES and absorber units. Individual components can be independently assembled and tested before integration into the larger module.  If one of the sub-modules malfunctions during initial testing, it can be set aside for repair or be replaced. The design will allow for the assembly of large arrays with close to 100\% of functional detectors.
The TESs will be read out using microwave multiplexing technique, which is \textcolor{black}{expected to be} well suited for space applications requiring large detector arrays, while \textcolor{black}{helping to} {minimize} system complexity and mass and {to provide} a high degree of ruggedness. 
In case of the SLEDGEHAMMER detectors, the thin silicon nitride membrane often broke when absorbers were attached, preventing the use of automated schemes for the fabrication of large arrays.  
%The new PRONGHORN design, placing the absorbers on Si carrier boards. 
In the PRONGHORN design, the absorbers are glued to a thick prong of Si, shown in Fig.\,\ref{f:unit} bottom right, giving a more robust support that should allow absorber placement to be automated.
The weak thermal link, previously accomplished through the delicate silicon nitride membranes, is now effected by
the Si legs that connect the TES and absorber on the prong to the rest of the Si substrate, as shown in Fig.\,\ref{f:unit} bottom left.
%In addition to its modularity, the new design can endure substantial physical handling and repeated attachment and detachment of the absorber. 
In addition to its modularity, the new design is expected to endure substantial physical handling. 
It is 
thus much better suited for automating the process 
of the absorber attachment and for achieving a very high yield of functional TES-absorber units.%\\[0.8ex]

To achieve high $\gamma$-ray detection efficiency, thick absorbers with high atomic numbers are required. 
The next section will discuss first results obtained with a prototype detector that uses 2\,mm thick Bi absorbers achieving a 511\,keV detection efficiency of 15\% for photoelectric effect events.

\subsection{First results obtained with thick Bi absorbers}
% \label{s:ph}

\begin{figure}[tb]
  \centering
  \includegraphics[width=\textwidth]{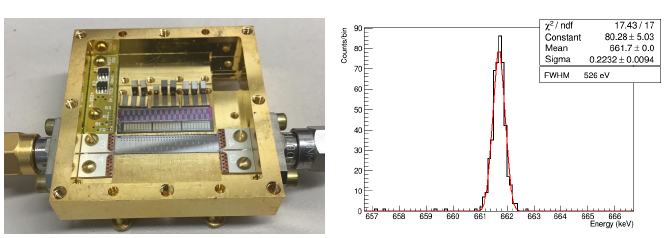}
  \caption{The left panel shows a prototype detector equipped with 3 PRONGHORN chips and 8 absorbers. The right panel shows the results from a test of the best channel, with a wafer diced Bi absorber, using a $^{137}$Cs source. We measure a 662\,keV energy resolution of 525\,eV FWHM.   
  }
  \label{f:511CD}
\end{figure}
The PRONGHORN approach allows us to separate the fabrication of TES-absorber units from the assembly of large modules.
We tested the feasibility of the approach by fabricating first detectors with Bi absorbers. Our set-up uses a detector box that can hold three PRONGHORN chips, each holding 3 absorbers. 
We fabricated 1.3$\times$1.3$\times$2 mm$^3$ absorbers made of Sn, Bi, and BiSn using three methods: Electrical Discharge Machining (EDM), 
casting (only Bi), and wafer dicing (Fig.\,\ref{f:unit}).
EDM and casting gave rough surfaces. EDM gave rough surfaces and spurs. The best results were obtained with wafer dicing. 
We equipped the PRONGHORN detector box with 3 cast BiSn absorbers and 3 wafer diced Bi absorbers (Fig.\,\ref{f:511CD} left panel). %All absorbers had a volume of 1.35$\times$1.35$\times$2\,mm$^2$. 
We obtained the best results for the Bi absorbers coupled to MoAu TESs. 
Tests with a $^{137}$Cs source revealed a 662\,keV energy resolution of 525\,eV FWHM (Fig.\,\ref{f:511CD}, right panel). \textcolor{black}{This result is preliminary and was obtained with a non-optimized detector and laboratory readout chain. Our ongoing improvements to the fabrication/assembly protocol and readout-noise control are expected to move the performance toward the 
$\sim 200$ eV goal. 
Our experimentally demonstrated energy resolution of 525 eV could be further improved by adopting a BiSn absorber. If BiSn becomes superconducting at the operating temperature ($\sim$ 100 mK), its heat capacity would be significantly reduced, thereby enhancing detector sensitivity. 
A detailed optimization study and noise budget will be presented in a follow-up publication.}

\section{Expected Results from one Long Duration Balloon Flight}
\label{s:outcome}
We estimate the sensitivity of the 511SM based on a galactic center 511\,keV flux of 10$^{-3}$ photons\,cm$^{-2}$ s$^{-1}$ \cite{2020ApJ...897...45S}, and the 511\,keV photoelectric cross section from XCOM
\cite{NIST_XCOM}. 
The background estimates are based on the appropriately scaled {\it XL-Calibur} background rates measured during the 2022 flight from Sweden to Canada  \cite{2021APh...12602529A,2023NIMPA104867975I}. These estimates are conservative as they have been derived from a high-latitude flight close to solar minimum. These background rates are consistent with an estimate from a detailed MGEANT simulations of a similar detector and shield assembly \cite{2023JATIS...9b4006S}.

\begin{figure}
\includegraphics[width=\textwidth]{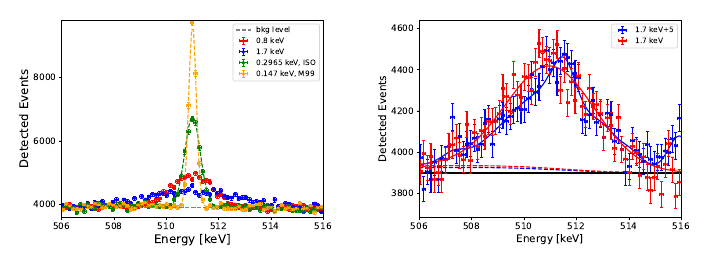} 
  \caption{Modeling the signal efficiency and passive and active background suppression, \textcolor{black}{assuming a detector FWHM energy resolution of 230 eV.} We estimate that 511SM would detect the 511\,keV galactic center emission with a signal-to-noise ratio of 35\,$\sigma$ for a 70-day super-pressure balloon flight from W\={a}naka, NZ. 
  Left: simulated results for different inherent widths (in $\sigma$, assuming a gaussian line profile) of the line, with ISO and M99 referring to the dark matter models from Ascasibar et al. 2006\cite{2006MNRAS.368.1695A}. 
  Right: simulated result for the case that 25\% of
  the flux comes from 5 bright regions assuming a frequency shift distribution consistent with the 
  4\,keV$\pm$0.9\,keV  FWHM line width measured by {\it COSI} \cite{2020ApJ...895...44K}. {The dashed curves display the continuum contribution from ortho-positronium.}
  The space-borne {\it COSI} mission will have an energy resolution requirement of 6\,keV FWHM, 26 times larger than that of 511SM, preventing it from measuring the line width with a small systematic error and from detecting the substructure shown here.   
   }
  \label{f:signal}
\end{figure}

The measurements from the balloon-borne {\it COSI} mission indicate that the   
511\,keV emission from the central 16$^{\circ}$-radius region around the galactic center has an inherent width of 
4\,keV$\pm$0.9\,keV FWHM \cite{2020ApJ...895...44K}.
Accounting for the 90\% duty cycle of the CMDs, 10\% signal loss at the entrance window, 5\% signal loss by the energy selection cut, the time the galactic center is above the horizon, and the time-of-day dependent atmospheric transmissivities, we estimate that 511SM will detect 15,400 511\,keV photons (signal rate: 2.96 mHz) on the 70-day super-pressure balloon flight. Assuming we integrate the signal over a 6.8\,keV window, the expected number of background events is 193,000 (background rate: 5.45 mHz).
We thus predict a line detection with 35\,$\sigma$ statistical significance.  
Fig.\,\ref{f:signal} shows that the 511SM would be able to measure the line width with exquisite precision, easily measuring the width and sub-structure of the line.
The right panel assumes that 25\% of the detected line flux comes from {5} bright astrophysical regions or sources, and that the annihilating electrons and positrons have eV-energies in the frames co-moving with these regions.

\section{Discussion}
\label{s:discussion}
In this paper, we describe the design of a balloon-borne 511SM, leveraging a TES array with thick high-Z metal absorbers and miniature cryogenic cooler. The TES detectors achieve unprecedented energy resolutions at $\gamma$-ray band, with a {target} FWHM of $\sim$200 eV at 511 keV. Assuming a 70-day super-pressure balloon flight, 511SM could measure the width and possible sub-structure of the galactic 511 keV emission line. Such observations would enable the identification of potential point-like astrophysical sources and provide novel insights on the positron sources of the galactic 511 keV emission, and help trace MeV dark matter in the galaxy.

The 511SM would be an ideal supplementary mission to the orbital {\it COSI} mission, which will be launched in 2027. The 511SM could provide follow-up observations with excellent spectral resolution on any bright sources observed by {\it COSI}, especially any point sources near the galactic center. Moreover, this mission could be a pathfinder for future MeV instruments using TES technologies as an alternative to the silicon	strip detectors, like {\it COSI} and {\it ComPair}.

The 511SM could be the precursor to a mission with better angular resolution. A possible follow-up mission could use coded mask imaging as the proposed {\it Galactic Annihilation Line Explorer (GALE)} experiment (which uses Cadmium Zinc Telluride as focal plane detectors) \cite{Moiseev2025}. In the coded mask system, incoming $\gamma$-ray photons are spatially modulated by a partially transparent mask (containing both transparent and opaque masks elements) before hitting the detector. The source image can be reconstructed through post-processing the observation \cite{Daniel2021}. The IBIS telescope on the {\it INTEGRAL} $\gamma$-ray space mission uses coded mask technology, and it has been used for searching point 511 keV sources near the galactic center \cite{2011A&A...531A..56D}. Alternatively, a follow-up mission like 511-CAM\cite{2023JATIS...9b4006S} could use $\gamma$-ray optics, possibly grazing incidence multilayer optics \cite{2023apra.prop...30S}, to focus the $\gamma$-rays onto the detector. The 2-D detector modules used in the 511SM design could be used for the coded mask version of the experiment as well. A 511-CAM mission would preferentially use stacked 2-D arrays to increase the detection efficiency to get close to 100\%. Indeed, the proposed 511-CAM mission plans for 8 layers of 2\,mm thick Bi absorbers giving a 
$>$90\% detection efficiency. 

\subsection* {Disclosures }
The authors declare that there are no financial interests, commercial affiliations, or other potential conflicts of interest that could have influenced the objectivity of this research or the writing of this paper.

\subsection* {Code and Data Availability}
The data utilized in this study are available from the authors upon request. The MASS software used to reduce the $^{137}$Cs spectrum is available in a Github repository (\url{https://github.com/usnistgov/mass}).

\subsection* {Acknowledgments}
We thank Carolyn Kierans, Casey DeRoo, Dustin Swarm, Danielle Gurgew, and Nicolas Barrière for fruitful discussions about 511 keV optics and science. 
KH and HK thank NASA for support under the grants 80NSSC22K1883 and 80NSSC21K1817, as well as the McDonnell Center for the Space Sciences for R\&D funds that enabled this research.

%%%%% References %%%%%

\bibliography{511SM}   % bibliography data in report.bib

@ARTICLE{1972ApJ...172L...1J,
       author = {{Johnson}, W.~N., III and {Harnden}, F.~R., Jr. and {Haymes}, R.~C.},
        title = "{The Spectrum of Low-Energy Gamma Radiation from the Galactic-Center Region.}",
      journal = {\apjl},
         year = 1972,
        month = feb,
       volume = {172},
        pages = {L1},
          doi = {10.1086/180878},
       adsurl = {https://ui.adsabs.harvard.edu/abs/1972ApJ...172L...1J}
}

@MISC{2023apra.prop...30S,
       author = {{Swarm}, Dustin},
        title = "{Designing and Fabricating Multilayer Optics to Enable a Future 511 keV Focusing Imager}",
 howpublished = {NASA Proposal ID. 23-APRA23-30},
         year = 2023,
        month = jan,
        pages = {30},
       adsurl = {https://ui.adsabs.harvard.edu/abs/2023apra.prop...30S}
}

@article{Daniel2021,
  author = {Daniel, G. and Limousin, O.},
  title = {{Extended sources reconstructions by means of coded mask aperture systems and deep learning algorithm}},
  journal = {Nuclear Instruments and Methods in Physics Research Section A: Accelerators, Spectrometers, Detectors and Associated Equipment},
  year = {2021},
  volume = {1012},
  doi = {10.1016/j.nima.2021.165600}
}

@online{NIST_XCOM,
  author = {Berger, M.J. and Hubbell, J.H. and Seltzer, S.M. and Chang, J. and Coursey, J.S. and Sukumar, R. and Zucker, D.S. and Olsen, K.},
  title = {{XCOM: Photon Cross Sections Database}},
  howpublished = {\url{https://www.nist.gov/pml/xcom-photon-cross-sections-database}},
  year = {2010},
  note = {Data content last updated November 2010; Page created September 17, 2009; Last updated August 14, 2024},
  publisher = {National Institute of Standards and Technology},
}

@ARTICLE{2025MNRAS.540L..34A,
       author = {{Awaki}, Hisamitsu and {Baring}, Matthew G. and {Bose}, Richard and {Braun}, Dana and {Casey}, Jacob and {Chun}, Sohee and {Galchenko}, Pavel and {Gau}, Ephraim and {Goya}, Kazuho and {Hakamata}, Tomohiro and {Hayashi}, Takayuki and {Heatwole}, Scott and {Hu}, Kun and {Imazawa}, Ryo and {Ishi}, Daiki and {Ishida}, Manabu and {Kislat}, Fabian and {Kiss}, M{\'o}zsi and {Klepper}, Kassi and {Krawczynski}, Henric and {Kuramoto}, Haruki and {Lanzi}, R. James and {Lisalda}, Lindsey and {Maeda}, Yoshitomo and {af Malmborg}, Filip and {Matsumoto}, Hironori and {Menon}, Shravan Vengalil and {Miyamoto}, Aiko and {Miyamoto}, Asca and {Miyazawa}, Takuya and {Murakami}, Kaito and {Nagao}, Azuki and {Okajima}, Takashi and {Pearce}, Mark and {Rauch}, Brian F. and {Rodriguez Cavero}, Nicole and {Shima}, Kohei and {Shirahama}, Kentaro and {Snow}, Carlton M. and {Spooner}, Sean and {Takahashi}, Hiromitsu and {Takatsuka}, Sayana and {Tamura}, Keisuke and {Tanaka}, Kojiro and {Uchida}, Yuusuke and {West}, Andrew Thomas and {Wulf}, Eric A. and {Yokota}, Masato and {Yoshimoto}, Marina},
        title = "{XL-Calibur measurements of polarized hard X-ray emission from the Crab}",
      journal = {\mnras},
         year = 2025,
        month = jun,
       volume = {540},
       number = {1},
        pages = {L34-L40},
          doi = {10.1093/mnrasl/slaf026},
archivePrefix = {arXiv},
       eprint = {2503.14307},
 primaryClass = {astro-ph.HE},
       adsurl = {https://ui.adsabs.harvard.edu/abs/2025MNRAS.540L..34A}
}

@ARTICLE{2018RScI...89l6105P,
       author = {{Plesha}, C.~E. and {Tonita}, E.~M. and {Kycia}, J.~B.},
        title = "{Note: Carbon fiber composite arrow shaft as cryogenic structural support material}",
      journal = {Review of Scientific Instruments},
         year = 2018,
        month = dec,
       volume = {89},
       number = {12},
          eid = {126105},
        pages = {126105},
          doi = {10.1063/1.5055899},
       adsurl = {https://ui.adsabs.harvard.edu/abs/2018RScI...89l6105P}
}

@ARTICLE{2023NIMPA104867975I,
       author = {{Iyer}, N.~K. and {Kiss}, M. and {Pearce}, M. and {Stana}, T. -A. and {Awaki}, H. and {Bose}, R.~G. and {Dasgupta}, A. and {De Geronimo}, G. and {Gau}, E. and {Hakamata}, T. and {Ishida}, M. and {Ishiwata}, K. and {Kamogawa}, W. and {Kislat}, F. and {Kitaguchi}, T. and {Krawczynski}, H. and {Lisalda}, L. and {Maeda}, Y. and {Matsumoto}, H. and {Miyamoto}, A. and {Miyazawa}, T. and {Mizuno}, T. and {Rauch}, B.~F. and {Cavero}, N. Rodriguez and {Sakamoto}, N. and {Sato}, J. and {Spooner}, S. and {Takahashi}, H. and {Takeo}, M. and {Tamagawa}, T. and {Uchida}, Y. and {West}, A.~T. and {Wimalasena}, K. and {Yoshimoto}, M.},
        title = "{The design and performance of the XL-Calibur anticoincidence shield}",
      journal = {Nuclear Instruments and Methods in Physics Research A},
         year = 2023,
        month = mar,
       volume = {1048},
          eid = {167975},
        pages = {167975},
          doi = {10.1016/j.nima.2022.167975},
archivePrefix = {arXiv},
       eprint = {2212.04139},
 primaryClass = {astro-ph.IM},
       adsurl = {https://ui.adsabs.harvard.edu/abs/2023NIMPA104867975I}
}

@misc{coax,
  title = {{Coax Co. LTD.}},
  howpublished = {\url{http://www.coax.co.jp/wcaxp/wp-content/uploads/2022/10/Superconducting_cable.pdf}},
  note = {Accessed on 1/14/2024}
}

@ARTICLE{2023arXiv230812362T,
       author = {{Tomsick}, John A. and {Boggs}, Steven E. and {Zoglauer}, Andreas and {Hartmann}, Dieter and {Ajello}, Marco and {Burns}, Eric and {Fryer}, Chris and {Karwin}, Chris and {Kierans}, Carolyn and {Lowell}, Alexander and {Malzac}, Julien and {Roberts}, Jarred and {Saint-Hilaire}, Pascal and {Shih}, Albert and {Siegert}, Thomas and {Sleator}, Clio and {Takahashi}, Tadayuki and {Tavecchio}, Fabrizio and {Wulf}, Eric and {Beechert}, Jacqueline and {Gulick}, Hannah and {Joens}, Alyson and {Lazar}, Hadar and {Neights}, Eliza and {Martinez Oliveros}, Juan Carlos and {Matsumoto}, Shigeki and {Melia}, Tom and {Yoneda}, Hiroki and {Amman}, Mark and {Bal}, Dhruv and {von Ballmoos}, Peter and {Bates}, Hugh and {B{\"o}ttcher}, Markus and {Bulgarelli}, Andrea and {Cavazzuti}, Elisabetta and {Chang}, Hsiang-Kuang and {Chen}, Claire and {Chu}, Che-Yen and {Ciabattoni}, Alex and {Costamante}, Luigi and {Dreyer}, Lente and {Fioretti}, Valentina and {Fenu}, Francesco and {Gallego}, Savitri and {Ghirlanda}, Giancarlo and {Grove}, Eric and {Huang}, Chien-You and {Jean}, Pierre and {Khatiya}, Nikita and {Kn{\"o}dlseder}, J{\"u}rgen and {Krause}, Martin and {Leising}, Mark and {Lewis}, Tiffany R. and {Lommler}, Jan Peter and {Marcotulli}, Lea and {Martinez-Castellanos}, Israel and {Mittal}, Saurabh and {Negro}, Michela and {Al Nussirat}, Samer and {Nakazawa}, Kazuhiro and {Oberlack}, Uwe and {Palmore}, David and {Panebianco}, Gabriele and {Parmiggiani}, Nicolo and {Parsotan}, Tyler and {Pike}, Sean N. and {Rogers}, Field and {Schutte}, Hester and {Sheng}, Yong and {Smale}, Alan P. and {Smith}, Jacob and {Trigg}, Aaron and {Venters}, Tonia and {Watanabe}, Yu and {Zhang}, Haocheng},
        title = "{The Compton Spectrometer and Imager}",
      journal = {arXiv e-prints},
         year = 2023,
        month = aug,
          eid = {arXiv:2308.12362},
        pages = {arXiv:2308.12362},
          doi = {10.48550/arXiv.2308.12362},
archivePrefix = {arXiv},
       eprint = {2308.12362},
 primaryClass = {astro-ph.HE},
       adsurl = {https://ui.adsabs.harvard.edu/abs/2023arXiv230812362T}
}

@ARTICLE{2022JLTP..209..337D,
       author = {{Devasia}, Archana M. and {Bandler}, Simon R. and {Ryu}, Kevin and {Stevenson}, Thomas R. and {Yoon}, Wonsik},
        title = "{Large-Scale Magnetic Microcalorimeter Arrays for the Lynx X-Ray Microcalorimeter}",
      journal = {Journal of Low Temperature Physics},
         year = 2022,
        month = nov,
       volume = {209},
       number = {3-4},
        pages = {337-345},
          doi = {10.1007/s10909-022-02767-z},
       adsurl = {https://ui.adsabs.harvard.edu/abs/2022JLTP..209..337D}
}

@INPROCEEDINGS{2022cosp...44.2316B,
       author = {{Barret}, Didier},
        title = "{The Athena X-ray Observatory}",
    booktitle = {44th COSPAR Scientific Assembly. Held 16-24 July},
         year = 2022,
       volume = {44},
        month = jul,
        pages = {2316},
       adsurl = {https://ui.adsabs.harvard.edu/abs/2022cosp...44.2316B}
}

@INPROCEEDINGS{2022SPIE12181E..1SI,
       author = {{Ishisaki}, Yoshitaka and {Kelley}, Richard L. and {Awaki}, Hisamitsu and {Balleza}, Jesus C. and {Barnstable}, Kim R. and {Bialas}, Thomas G. and {Boissay-Malaquin}, Rozenn and {Brown}, Gregory V. and {Canavan}, Edgar R. and {Cumbee}, Renata S. and {Carnahan}, Timothy M. and {Chiao}, Meng P. and {Comber}, Brian J. and {Costantini}, Elisa and {den Herder}, Jan-Willem and {Dercksen}, Johannes and {de Vries}, Cor P. and {DiPirro}, Michael J. and {Eckart}, Megan E. and {Ezoe}, Yuichiro and {Ferrigno}, Carlo and {Fujimoto}, Ryuichi and {Gorter}, Nathalie and {Graham}, Steven M. and {Grim}, Martin and {Hartz}, Leslie S. and {Hayakawa}, Ryota and {Hayashi}, Takayuki and {Hell}, Natalie and {Hoshino}, Akio and {Ichinohe}, Yuto and {Ishida}, Manabu and {Ishikawa}, Kumi and {James}, Bryan L. and {Kenyon}, Steven J. and {Kilbourne}, Caroline A. and {Kimball}, Mark O. and {Kitamoto}, Shunji and {Leutenegger}, Maurice A. and {Maeda}, Yoshitomo and {McCammon}, Dan and {Miko}, Joseph J. and {Mizumoto}, Misaki and {Okajima}, Takashi and {Okamoto}, Atsushi and {Paltani}, Stephane and {Porter}, Frederick S. and {Sato}, Kosuke and {Sato}, Toshiki and {Sawada}, Makoto and {Shinozaki}, Keisuke and {Shipman}, Russell and {Shirron}, Peter J. and {Sneiderman}, Gary A. and {Soong}, Yang and {Szymkiewicz}, Richard and {Szymkowiak}, Andrew E. and {Takei}, Yoh and {Tamura}, Keisuke and {Tsujimoto}, Masahiro and {Uchida}, Yuusuke and {Wasserzug}, Stephen and {Witthoeft}, Michael C. and {Wolfs}, Rob and {Yamada}, Shinya and {Yasuda}, Susumu},
        title = "{Status of resolve instrument onboard X-Ray Imaging and Spectroscopy Mission (XRISM)}",
    booktitle = {Space Telescopes and Instrumentation 2022: Ultraviolet to Gamma Ray},
         year = 2022,
       editor = {{den Herder}, Jan-Willem A. and {Nikzad}, Shouleh and {Nakazawa}, Kazuhiro},
       series = {Society of Photo-Optical Instrumentation Engineers (SPIE) Conference Series},
       volume = {12181},
        month = aug,
          eid = {121811S},
        pages = {121811S},
          doi = {10.1117/12.2630654},
       adsurl = {https://ui.adsabs.harvard.edu/abs/2022SPIE12181E..1SI}
}

@ARTICLE{2018MNRAS.479.2229C,
       author = {{Chan}, Man Ho and {Leung}, Chung Hei},
        title = "{Constraining dark matter by the 511 keV line}",
      journal = {\mnras},
         year = 2018,
        month = sep,
       volume = {479},
       number = {2},
        pages = {2229-2234},
          doi = {10.1093/mnras/sty1583},
archivePrefix = {arXiv},
       eprint = {1806.07102},
 primaryClass = {astro-ph.HE},
       adsurl = {https://ui.adsabs.harvard.edu/abs/2018MNRAS.479.2229C}
}

@ARTICLE{noroozian_etal_2013,
   author = {{Noroozian}, O. and {Mates}, J.~A.~B. and {Bennett}, D.~A. and 
	{Brevik}, J.~A. and {Fowler}, J.~W. and {Gao}, J. and {Hilton}, G.~C. and 
	{Horansky}, R.~D. and {Irwin}, K.~D. and {Kang}, Z. and {Schmidt}, D.~R. and 
	{Vale}, L.~R. and {Ullom}, J.~N.},
    title = "{High-resolution gamma-ray spectroscopy with a microwave-multiplexed transition-edge sensor array}",
  journal = {Appl. Phys. Lett.},
archivePrefix = "arXiv",
   eprint = {1310.7287},
 primaryClass = "physics.ins-det",
     year = 2013,
    month = nov,
   volume = 103,
   number = 20,
      eid = {202602},
    pages = {202602},
      doi = {10.1063/1.4829156},
   adsurl = {http://adsabs.harvard.edu/abs/2013ApPhL.103t2602N}
}

@ARTICLE{2010ApJ...720.1772B,
       author = {{Bouchet}, L. and {Roques}, J.~P. and {Jourdain}, E.},
        title = "{On the Morphology of the Electron-Positron Annihilation Emission as Seen by Spi/integral}",
      journal = {\apj},
         year = 2010,
        month = sep,
       volume = {720},
       number = {2},
        pages = {1772-1780},
          doi = {10.1088/0004-637X/720/2/1772},
archivePrefix = {arXiv},
       eprint = {1007.4753},
 primaryClass = {astro-ph.HE},
       adsurl = {https://ui.adsabs.harvard.edu/abs/2010ApJ...720.1772B}
}

@ARTICLE{2005MNRAS.357.1377C,
       author = {{Churazov}, E. and {Sunyaev}, R. and {Sazonov}, S. and {Revnivtsev}, M. and {Varshalovich}, D.},
        title = "{Positron annihilation spectrum from the Galactic Centre region observed by SPI/INTEGRAL}",
      journal = {\mnras},
         year = 2005,
        month = mar,
       volume = {357},
       number = {4},
        pages = {1377-1386},
          doi = {10.1111/j.1365-2966.2005.08757.x},
archivePrefix = {arXiv},
       eprint = {astro-ph/0411351},
 primaryClass = {astro-ph},
       adsurl = {https://ui.adsabs.harvard.edu/abs/2005MNRAS.357.1377C}
}

@ARTICLE{2011MNRAS.411.1727C,
       author = {{Churazov}, E. and {Sazonov}, S. and {Tsygankov}, S. and {Sunyaev}, R. and {Varshalovich}, D.},
        title = "{Positron annihilation spectrum from the Galactic Centre region observed by SPI/INTEGRAL revisited: annihilation in a cooling ISM?}",
      journal = {\mnras},
         year = 2011,
        month = mar,
       volume = {411},
       number = {3},
        pages = {1727-1743},
          doi = {10.1111/j.1365-2966.2010.17804.x},
archivePrefix = {arXiv},
       eprint = {1010.0864},
 primaryClass = {astro-ph.HE},
       adsurl = {https://ui.adsabs.harvard.edu/abs/2011MNRAS.411.1727C}
}

@ARTICLE{2005A&A...441..513K,
       author = {{Kn{\"o}dlseder}, J. and {Jean}, P. and {Lonjou}, V. and {Weidenspointner}, G. and {Guessoum}, N. and {Gillard}, W. and {Skinner}, G. and {von Ballmoos}, P. and {Vedrenne}, G. and {Roques}, J. -P. and {Schanne}, S. and {Teegarden}, B. and {Sch{\"o}nfelder}, V. and {Winkler}, C.},
        title = "{The all-sky distribution of 511 keV electron-positron annihilation emission}",
      journal = {\aap},
         year = 2005,
        month = oct,
       volume = {441},
       number = {2},
        pages = {513-532},
          doi = {10.1051/0004-6361:20042063},
archivePrefix = {arXiv},
       eprint = {astro-ph/0506026},
 primaryClass = {astro-ph},
       adsurl = {https://ui.adsabs.harvard.edu/abs/2005A&A...441..513K}
}

@ARTICLE{1997ApJ...491..725P,
       author = {{Purcell}, W.~R. and {Cheng}, L. -X. and {Dixon}, D.~D. and {Kinzer}, R.~L. and {Kurfess}, J.~D. and {Leventhal}, M. and {Saunders}, M.~A. and {Skibo}, J.~G. and {Smith}, D.~M. and {Tueller}, J.},
        title = "{OSSE Mapping of Galactic 511 keV Positron Annihilation Line Emission}",
      journal = {\apj},
         year = 1997,
        month = dec,
       volume = {491},
       number = {2},
        pages = {725-748},
          doi = {10.1086/304994},
       adsurl = {https://ui.adsabs.harvard.edu/abs/1997ApJ...491..725P}
}

@ARTICLE{2008Natur.451..159W,
       author = {{Weidenspointner}, Georg and {Skinner}, Gerry and {Jean}, Pierre and {Kn{\"o}dlseder}, J{\"u}rgen and {von Ballmoos}, Peter and {Bignami}, Giovanni and {Diehl}, Roland and {Strong}, Andrew W. and {Cordier}, Bertrand and {Schanne}, St{\'e}phane and {Winkler}, Christoph},
        title = "{An asymmetric distribution of positrons in the Galactic disk revealed by {\ensuremath{\gamma}}-rays}",
      journal = {\nat},
         year = 2008,
        month = jan,
       volume = {451},
       number = {7175},
        pages = {159-162},
          doi = {10.1038/nature06490},
       adsurl = {https://ui.adsabs.harvard.edu/abs/2008Natur.451..159W}
}

@ARTICLE{2020ApJ...897...45S,
       author = {{Siegert}, Thomas and {Boggs}, Steven E. and {Tomsick}, John A. and {Zoglauer}, Andreas C. and {Kierans}, Carolyn A. and {Sleator}, Clio C. and {Beechert}, Jacqueline and {Brandt}, Theresa J. and {Jean}, Pierre and {Lazar}, Hadar and {Lowell}, Alex W. and {Roberts}, Jarred M. and {Ballmoos}, Peter von},
        title = "{Imaging the 511 keV Positron Annihilation Sky with COSI}",
      journal = {\apj},
         year = 2020,
        month = jul,
       volume = {897},
       number = {1},
          eid = {45},
        pages = {45},
          doi = {10.3847/1538-4357/ab9607},
archivePrefix = {arXiv},
       eprint = {2005.10950},
 primaryClass = {astro-ph.HE},
       adsurl = {https://ui.adsabs.harvard.edu/abs/2020ApJ...897...45S}
}

@ARTICLE{2011RvMP...83.1001P,
       author = {{Prantzos}, N. and {Boehm}, C. and {Bykov}, A.~M. and {Diehl}, R. and {Ferri{\`e}re}, K. and {Guessoum}, N. and {Jean}, P. and {Knoedlseder}, J. and {Marcowith}, A. and {Moskalenko}, I.~V. and {Strong}, A. and {Weidenspointner}, G.},
        title = "{The 511 keV emission from positron annihilation in the Galaxy}",
      journal = {Reviews of Modern Physics},
         year = 2011,
        month = jul,
       volume = {83},
       number = {3},
        pages = {1001-1056},
          doi = {10.1103/RevModPhys.83.1001},
archivePrefix = {arXiv},
       eprint = {1009.4620},
 primaryClass = {astro-ph.HE},
       adsurl = {https://ui.adsabs.harvard.edu/abs/2011RvMP...83.1001P}
}

@ARTICLE{2011A&A...531A..56D,
       author = {{De Cesare}, G.},
        title = "{Searching for the 511 keV annihilation line from galactic compact objects with the IBIS gamma ray telescope}",
      journal = {\aap},
         year = 2011,
        month = jul,
       volume = {531},
          eid = {A56},
        pages = {A56},
          doi = {10.1051/0004-6361/201116516},
archivePrefix = {arXiv},
       eprint = {1105.0367},
 primaryClass = {astro-ph.HE},
       adsurl = {https://ui.adsabs.harvard.edu/abs/2011A&A...531A..56D}
}

@ARTICLE{2023JATIS...9b4006S,
       author = {{Shirazi}, Farzane and {Gau}, Ephraim and {Hossen}, Md. Arman and {Becker}, Daniel and {Schmidt}, Daniel and {Swetz}, Daniel and {Bennett}, Douglas and {Braun}, Dana and {Kislat}, Fabian and {Gard}, Johnathon and {Mates}, John and {Weber}, Joel and {Rodriguez Cavero}, Nicole and {Chun}, Sohee and {Lisalda}, Lindsey and {West}, Andrew and {Dev}, Bhupal and {Ferrer}, Francesc and {Bose}, Richard and {Ullom}, Joel and {Krawczynski}, Henric},
        title = "{511-CAM mission: a pointed 511 keV gamma-ray telescope with a focal plane detector made of stacked transition edge sensor microcalorimeter arrays}",
      journal = {Journal of Astronomical Telescopes, Instruments, and Systems},
         year = 2023,
        month = apr,
       volume = {9},
          eid = {024006},
        pages = {024006},
          doi = {10.1117/1.JATIS.9.2.024006},
archivePrefix = {arXiv},
       eprint = {2206.14652},
 primaryClass = {astro-ph.IM},
       adsurl = {https://ui.adsabs.harvard.edu/abs/2023JATIS...9b4006S}
}

@ARTICLE{2018A&A...616A..83V,
       author = {{Valenti}, E. and {Zoccali}, M. and {Mucciarelli}, A. and {Gonzalez}, O.~A. and {Surot}, F. and {Minniti}, D. and {Rejkuba}, M. and {Pasquini}, L. and {Fiorentino}, G. and {Bono}, G. and {Rich}, R.~M. and {Soto}, M.},
        title = "{The central velocity dispersion of the Milky Way bulge}",
      journal = {\aap},
         year = 2018,
        month = aug,
       volume = {616},
          eid = {A83},
        pages = {A83},
          doi = {10.1051/0004-6361/201832905},
archivePrefix = {arXiv},
       eprint = {1805.00275},
 primaryClass = {astro-ph.GA},
       adsurl = {https://ui.adsabs.harvard.edu/abs/2018A&A...616A..83V}
}

@INPROCEEDINGS{2019SPIE11118E..0KS,
       author = {{Schwartz}, Daniel A. and {Vikhlinin}, Alexey and {Tananbaum}, Harvey and {Freeman}, Mark and {Tremblay}, Grant and {Schwartz}, Eric D. and {Gaskin}, Jessica A. and {Swartz}, Douglas and {Gelmis}, Karen and {McCarley}, Kevin S. and {Dominguez}, Alexandra},
        title = "{The Lynx X-ray Observatory: revealing the invisible universe}",
    booktitle = {UV, X-Ray, and Gamma-Ray Space Instrumentation for Astronomy XXI},
         year = 2019,
       editor = {{Siegmund}, Oswald H.},
       series = {Society of Photo-Optical Instrumentation Engineers (SPIE) Conference Series},
       volume = {11118},
        month = sep,
          eid = {111180K},
        pages = {111180K},
          doi = {10.1117/12.2533637},
       adsurl = {https://ui.adsabs.harvard.edu/abs/2019SPIE11118E..0KS}
}

@ARTICLE{2022arXiv221109827K,
       author = {{Kraft}, Ralph and {Markevitch}, Maxim and {Kilbourne}, Caroline and {Adams}, Joseph S. and {Akamatsu}, Hiroki and {Ayromlou}, Mohammadreza and {Bandler}, Simon R. and {Barbera}, Marco and {Bennett}, Douglas A. and {Bhardwaj}, Anil and {Biffi}, Veronica and {Bodewits}, Dennis and {Bogdan}, Akos and {Bonamente}, Massimiliano and {Borgani}, Stefano and {Branduardi-Raymont}, Graziella and {Bregman}, Joel N. and {Burchett}, Joseph N. and {Cann}, Jenna and {Carter}, Jenny and {Chakraborty}, Priyanka and {Churazov}, Eugene and {Crain}, Robert A. and {Cumbee}, Renata and {Dave}, Romeel and {DiPirro}, Michael and {Dolag}, Klaus and {Bertrand Doriese}, W. and {Drake}, Jeremy and {Dunn}, William and {Eckart}, Megan and {Eckert}, Dominique and {Ettori}, Stefano and {Forman}, William and {Galeazzi}, Massimiliano and {Gall}, Amy and {Gatuzz}, Efrain and {Hell}, Natalie and {Hodges-Kluck}, Edmund and {Jackman}, Caitriona and {Jahromi}, Amir and {Jennings}, Fred and {Jones}, Christine and {Kaaret}, Philip and {Kavanagh}, Patrick J. and {Kelley}, Richard L. and {Khabibullin}, Ildar and {Kim}, Chang-Goo and {Koutroumpa}, Dimitra and {Kovacs}, Orsolya and {Kuntz}, K.~D. and {Lau}, Erwin and {Lee}, Shiu-Hang and {Leutenegger}, Maurice and {Lin}, Sheng-Chieh and {Lisse}, Carey and {Lo Cicero}, Ugo and {Lovisari}, Lorenzo and {McCammon}, Dan and {McEntee}, Sean and {Mernier}, Francois and {Miller}, Eric D. and {Nagai}, Daisuke and {Negro}, Michela and {Nelson}, Dylan and {Ness}, Jan-Uwe and {Nulsen}, Paul and {Ogorzalek}, Anna and {Oppenheimer}, Benjamin D. and {Oskinova}, Lidia and {Patnaude}, Daniel and {Pfeifle}, Ryan W. and {Pillepich}, Annalisa and {Plucinsky}, Paul and {Pooley}, David and {Porter}, Frederick S. and {Randall}, Scott and {Rasia}, Elena and {Raymond}, John and {Ruszkowski}, Mateusz and {Sakai}, Kazuhiro and {Sarkar}, Arnab and {Sasaki}, Manami and {Sato}, Kosuke and {Schellenberger}, Gerrit and {Schaye}, Joop and {Simionescu}, Aurora and {Smith}, Stephen J. and {Steiner}, James F. and {Stern}, Jonathan and {Su}, Yuanyuan and {Sun}, Ming and {Tremblay}, Grant and {Truong}, Nhut and {Tutt}, James and {Ursino}, Eugenio and {Veilleux}, Sylvain and {Vikhlinin}, Alexey and {Vladutescu-Zopp}, Stephan and {Vogelsberger}, Mark and {Walker}, Stephen A. and {Weaver}, Kimberly and {Weigt}, Dale M. and {Werk}, Jessica and {Werner}, Norbert and {Wolk}, Scott J. and {Zhang}, Congyao and {Zhang}, William W. and {Zhuravleva}, Irina and {ZuHone}, John},
        title = "{Line Emission Mapper (LEM): Probing the physics of cosmic ecosystems}",
      journal = {arXiv e-prints},
         year = 2022,
        month = nov,
          eid = {arXiv:2211.09827},
        pages = {arXiv:2211.09827},
          doi = {10.48550/arXiv.2211.09827},
archivePrefix = {arXiv},
       eprint = {2211.09827},
 primaryClass = {astro-ph.IM},
       adsurl = {https://ui.adsabs.harvard.edu/abs/2022arXiv221109827K}
}

@ARTICLE{2020ApJ...895...44K,
       author = {{Kierans}, C.~A. and {Boggs}, S.~E. and {Zoglauer}, A. and {Lowell}, A.~W. and {Sleator}, C. and {Beechert}, J. and {Brandt}, T.~J. and {Jean}, P. and {Lazar}, H. and {Roberts}, J. and {Siegert}, T. and {Tomsick}, J.~A. and {Ballmoos}, P. von},
        title = "{Detection of the 511 keV Galactic Positron Annihilation Line with COSI}",
      journal = {\apj},
         year = 2020,
        month = may,
       volume = {895},
       number = {1},
          eid = {44},
        pages = {44},
          doi = {10.3847/1538-4357/ab89a9},
archivePrefix = {arXiv},
       eprint = {1912.00110},
 primaryClass = {astro-ph.HE},
       adsurl = {https://ui.adsabs.harvard.edu/abs/2020ApJ...895...44K}
}

@ARTICLE{mates_etal_2017,
   author = {{Mates}, J.~A.~B. and {Becker}, D.~T. and {Bennett}, D.~A. and 
	{Dober}, B.~J. and {Gard}, J.~D. and {Hays-Wehle}, J.~P. and 
	{Fowler}, J.~W. and {Hilton}, G.~C. and {Reintsema}, C.~D. and 
	{Schmidt}, D.~R. and {Swetz}, D.~S. and {Vale}, L.~R. and {Ullom}, J.~N.
	},
    title = "{Simultaneous readout of 128 X-ray and gamma-ray transition-edge microcalorimeters using microwave SQUID multiplexing}",
  journal = {Appl. Phys. Lett.},
     year = 2017,
    month = aug,
   volume = 111,
   number = 6,
      eid = {062601},
    pages = {062601},
      doi = {10.1063/1.4986222},
   adsurl = {http://adsabs.harvard.edu/abs/2017ApPhL.111f2601M}
}

@ARTICLE{2012RScI...83i3113B,
   author = {{Bennett}, D.~A. and {Horansky}, R.~D. and {Schmidt}, D.~R. and 
	{Hoover}, A.~S. and {Winkler}, R. and {Alpert}, B.~K. and {Beall}, J.~A. and 
	{Doriese}, W.~B. and {Fowler}, J.~W. and {Fitzgerald}, C.~P. and 
	{Hilton}, G.~C. and {Irwin}, K.~D. and {Kotsubo}, V. and {Mates}, J.~A.~B. and 
	{O'Neil}, G.~C. and {Rabin}, M.~W. and {Reintsema}, C.~D. and 
	{Schima}, F.~J. and {Swetz}, D.~S. and {Vale}, L.~R. and {Ullom}, J.~N.
	},
    title = "{A high resolution gamma-ray spectrometer based on superconducting microcalorimeters}",
  journal = {Rev. Sci. Instrum.},
     year = 2012,
    month = sep,
   volume = 83,
   number = 9,
      eid = {093113-093113-14},
    pages = {093113-093113-14},
      doi = {10.1063/1.4754630},
   adsurl = {http://adsabs.harvard.edu/abs/2012RScI...83i3113B}
}

@techreport{stuchlik,
    author = {{Stuchlik}, D.~W. and {Heatwole}, S.},
    title = "{Celestial Attitude Reference and Determination System (CARDS) Daytime Star Tracker}",
    year = {2018},
    institution = {NASA},
    url = {https://ntrs.nasa.gov/citations/20180007599}
}

@ARTICLE{2021APh...12602529A,
       author = {{Abarr}, Q. and {Awaki}, H. and {Baring}, M.~G. and {Bose}, R. and {De Geronimo}, G. and {Dowkontt}, P. and {Errando}, M. and {Guarino}, V. and {Hattori}, K. and {Hayashida}, K. and {Imazato}, F. and {Ishida}, M. and {Iyer}, N.~K. and {Kislat}, F. and {Kiss}, M. and {Kitaguchi}, T. and {Krawczynski}, H. and {Lisalda}, L. and {Matake}, H. and {Maeda}, Y. and {Matsumoto}, H. and {Mineta}, T. and {Miyazawa}, T. and {Mizuno}, T. and {Okajima}, T. and {Pearce}, M. and {Rauch}, B.~F. and {Ryde}, F. and {Shreves}, C. and {Spooner}, S. and {Stana}, T. -A. and {Takahashi}, H. and {Takeo}, M. and {Tamagawa}, T. and {Tamura}, K. and {Tsunemi}, H. and {Uchida}, N. and {Uchida}, Y. and {West}, A.~T. and {Wulf}, E.~A. and {Yamamoto}, R.},
        title = "{XL-Calibur - a second-generation balloon-borne hard X-ray polarimetry mission}",
      journal = {Astroparticle Physics},
         year = 2021,
        month = mar,
       volume = {126},
          eid = {102529},
        pages = {102529},
          doi = {10.1016/j.astropartphys.2020.102529},
archivePrefix = {arXiv},
       eprint = {2010.10608},
 primaryClass = {astro-ph.IM},
       adsurl = {https://ui.adsabs.harvard.edu/abs/2021APh...12602529A}
}

@INPROCEEDINGS{2020SPIE11445E..7AL,
       author = {{Lowe}, Ian and {Coppi}, Gabriele and {Ade}, Peter A.~R. and {Ashton}, Peter C. and {Austermann}, Jason E. and {Beall}, James and {Clark}, Susan and {Cox}, Erin G. and {Devlin}, Mark J. and {Dicker}, Simon and {Dober}, Bradley J. and {Fanfani}, Valentina and {Fissel}, Laura M. and {Galitzki}, Nicholas and {Gao}, Jiangsong and {Hensley}, Brandon and {Hubmayr}, Johannes and {Li}, Steven and {Li}, Zhi-yun and {Lourie}, Nathan P. and {Martin}, Peter G. and {Mauskopf}, Philip and {Nati}, Federico and {Novak}, Giles and {Pisano}, Giampaolo and {Romualdez}, Javier L. and {Sinclair}, Adrian and {Soler}, Juan D. and {Tucker}, Carole and {Vissers}, Michael and {Wheeler}, Jordan and {Williams}, Paul A. and {Zannoni}, Mario},
        title = "{The Balloon-borne Large Aperture Submillimeter Telescope Observatory}",
    booktitle = {Ground-based and Airborne Telescopes VIII},
         year = 2020,
       editor = {{Marshall}, Heather K. and {Spyromilio}, Jason and {Usuda}, Tomonori},
       series = {Society of Photo-Optical Instrumentation Engineers (SPIE) Conference Series},
       volume = {11445},
        month = dec,
          eid = {114457A},
        pages = {114457A},
          doi = {10.1117/12.2576146},
archivePrefix = {arXiv},
       eprint = {2012.01376},
 primaryClass = {astro-ph.IM},
       adsurl = {https://ui.adsabs.harvard.edu/abs/2020SPIE11445E..7AL}
}

@ARTICLE{2014JAI.....340001G,
       author = {{Galitzki}, Nicholas and {Ade}, Peter A.~R. and {Angil{\`e}}, Francesco E. and {Ashton}, Peter and {Beall}, James A. and {Becker}, Dan and {Bradford}, Kristi J. and {Che}, George and {Cho}, Hsiao-Mei and {Devlin}, Mark J. and {Dober}, Bradley J. and {Fissel}, Laura M. and {Fukui}, Yasuo and {Gao}, Jiansong and {Groppi}, Christopher E. and {Hillbrand}, Seth and {Hilton}, Gene C. and {Hubmayr}, Johannes and {Irwin}, Kent D. and {Klein}, Jeffrey and {van Lanen}, Jeff and {Li}, Dale and {Li}, Zhi-Yun and {Lourie}, Nathan P. and {Mani}, Hamdi and {Martin}, Peter G. and {Mauskopf}, Philip and {Nakamura}, Fumitaka and {Novak}, Giles and {Pappas}, David P. and {Pascale}, Enzo and {Pisano}, Giampaolo and {Santos}, Fabio P. and {Savini}, Giorgio and {Scott}, Douglas and {Stanchfield}, Sara and {Tucker}, Carole and {Ullom}, Joel N. and {Underhill}, Matthew and {Vissers}, Michael R. and {Ward-Thompson}, Derek},
        title = "{The Next Generation BLAST Experiment}",
      journal = {Journal of Astronomical Instrumentation},
         year = 2014,
        month = nov,
       volume = {3},
       number = {2},
          eid = {1440001},
        pages = {1440001},
          doi = {10.1142/S2251171714400017},
archivePrefix = {arXiv},
       eprint = {1409.7084},
 primaryClass = {astro-ph.IM},
       adsurl = {https://ui.adsabs.harvard.edu/abs/2014JAI.....340001G}
}

@ARTICLE{gudmundsson_etal_2015,
   author = {{Gudmundsson}, J.~E. and {Ade}, P.~A.~R. and {Amiri}, M. and 
	{Benton}, S.~J. and {Bock}, J.~J. and {Bond}, J.~R. and {Bryan}, S.~A. and 
	{Chiang}, H.~C. and {Contaldi}, C.~R. and {Crill}, B.~P. and 
	{Dore}, O. and {Filippini}, J.~P. and {Fraisse}, A.~A. and {Gambrel}, A. and 
	{Gandilo}, N.~N. and {Hasselfield}, M. and {Halpern}, M. and 
	{Hilton}, G. and {Holmes}, W. and {Hristov}, V.~V. and {Irwin}, K.~D. and 
	{Jones}, W.~C. and {Kermish}, Z. and {MacTavish}, C.~J. and 
	{Mason}, P.~V. and {Megerian}, K. and {Moncelsi}, L. and {Montroy}, T.~E. and 
	{Morford}, T.~A. and {Nagy}, J.~M. and {Netterfield}, C.~B. and 
	{Rahlin}, A.~S. and {Reintsema}, C.~D. and {Ruhl}, J.~E. and 
	{Runyan}, M.~C. and {Shariff}, J.~A. and {Soler}, J.~D. and 
	{Trangsrud}, A. and {Tucker}, C. and {Tucker}, R.~S. and {Turner}, A.~D. and 
	{Wiebe}, D.~V. and {Young}, E. and {Spider Collaboration} and 
	{Abe}, P.~A.~R. and {Amiri}, M. and {Benton}, S.~J. and {Bock}, J.~J. and 
	{Bond}, J.~R. and {Bryan}, S.~A. and {Chiang}, H.~C. and {Contaldi}, C.~R. and 
	{Crill}, B.~P. and {Dore}, O. and {Filippini}, J.~P. and {Fraisse}, A.~A. and 
	{Gambrel}, A. and {Gandilo}, N.~N. and {Hasselfield}, M. and 
	{Halpern}, M. and {Hilton}, G. and {Holmes}, W. and {Hristov}, V.~V. and 
	{Irwin}, K.~D. and {Jones}, W.~C. and {Kermish}, Z. and {MacTavish}, C.~J. and 
	{Mason}, P.~V. and {Megerian}, K. and {Moncelsi}, L. and {Montroy}, T.~E. and 
	{Morford}, T.~A. and {Nagy}, J.~M. and {Netterfield}, C.~B. and 
	{Rahlin}, A.~S. and {Reintsema}, C.~D. and {Ruhl}, J.~E. and 
	{Runyan}, M.~C. and {Shariff}, J.~A. and {Soler}, J.~D. and 
	{Trangsrud}, A. and {Tucker}, C. and {Tucker}, R.~S. and {Turner}, A.~D. and 
	{Wiebe}, D.~V. and {Young}, E.},
    title = "{The thermal design, characterization, and performance of the SPIDER long-duration balloon cryostat}",
  journal = {Cryogenics},
archivePrefix = "arXiv",
   eprint = {1506.06953},
 primaryClass = "astro-ph.IM",
     year = 2015,
    month = dec,
   volume = 72,
    pages = {65-76},
      doi = {10.1016/j.cryogenics.2015.09.002},
   adsurl = {http://adsabs.harvard.edu/abs/2015Cryo...72...65G}
}

@ARTICLE{2004PhRvL..92j1301B,
       author = {{Boehm}, C{\'e}line and {Hooper}, Dan and {Silk}, Joseph and {Casse}, Michel and {Paul}, Jacques},
        title = "{MeV Dark Matter: Has It Been Detected?}",
      journal = {\prl},
         year = 2004,
        month = mar,
       volume = {92},
       number = {10},
          eid = {101301},
        pages = {101301},
          doi = {10.1103/PhysRevLett.92.101301},
archivePrefix = {arXiv},
       eprint = {astro-ph/0309686},
 primaryClass = {astro-ph},
       adsurl = {https://ui.adsabs.harvard.edu/abs/2004PhRvL..92j1301B}
}

@ARTICLE{2004PhRvD..70f3506H,
       author = {{Hooper}, Dan and {Wang}, Lian-Tao},
        title = "{Possible evidence for axino dark matter in the galactic bulge}",
      journal = {\prd},
         year = 2004,
        month = sep,
       volume = {70},
       number = {6},
          eid = {063506},
        pages = {063506},
          doi = {10.1103/PhysRevD.70.063506},
archivePrefix = {arXiv},
       eprint = {hep-ph/0402220},
 primaryClass = {astro-ph},
       adsurl = {https://ui.adsabs.harvard.edu/abs/2004PhRvD..70f3506H}
}

@ARTICLE{2007PhRvD..76h3519F,
       author = {{Finkbeiner}, Douglas P. and {Weiner}, Neal},
        title = "{Exciting dark matter and the INTEGRAL/SPI 511keV signal}",
      journal = {\prd},
         year = 2007,
        month = oct,
       volume = {76},
       number = {8},
          eid = {083519},
        pages = {083519},
          doi = {10.1103/PhysRevD.76.083519},
archivePrefix = {arXiv},
       eprint = {astro-ph/0702587},
 primaryClass = {astro-ph},
       adsurl = {https://ui.adsabs.harvard.edu/abs/2007PhRvD..76h3519F}
}

@ARTICLE{2006A&A...457..753G,
       author = {{Guessoum}, N. and {Jean}, P. and {Prantzos}, N.},
        title = "{Microquasars as sources of positron annihilation radiation}",
      journal = {\aap},
         year = 2006,
        month = oct,
       volume = {457},
       number = {3},
        pages = {753-762},
          doi = {10.1051/0004-6361:20065240},
archivePrefix = {arXiv},
       eprint = {astro-ph/0607296},
 primaryClass = {astro-ph},
       adsurl = {https://ui.adsabs.harvard.edu/abs/2006A&A...457..753G}
}

@ARTICLE{2009ApJ...698..350H,
       author = {{Higdon}, J.~C. and {Lingenfelter}, R.~E. and {Rothschild}, R.~E.},
        title = "{The Galactic Positron Annihilation Radiation and the Propagation of Positrons in the Interstellar Medium}",
      journal = {\apj},
         year = 2009,
        month = jun,
       volume = {698},
       number = {1},
        pages = {350-379},
          doi = {10.1088/0004-637X/698/1/350},
archivePrefix = {arXiv},
       eprint = {0711.3008},
 primaryClass = {astro-ph},
       adsurl = {https://ui.adsabs.harvard.edu/abs/2009ApJ...698..350H}
}

@ARTICLE{2014A&A...564A.108A,
       author = {{Alexis}, A. and {Jean}, P. and {Martin}, P. and {Ferri{\`e}re}, K.},
        title = "{Monte Carlo modelling of the propagation and annihilation of nucleosynthesis positrons in the Galaxy}",
      journal = {\aap},
         year = 2014,
        month = apr,
       volume = {564},
          eid = {A108},
        pages = {A108},
          doi = {10.1051/0004-6361/201322393},
archivePrefix = {arXiv},
       eprint = {1402.6110},
 primaryClass = {astro-ph.HE},
       adsurl = {https://ui.adsabs.harvard.edu/abs/2014A&A...564A.108A}
}

@ARTICLE{2006MNRAS.368.1695A,
       author = {{Ascasibar}, Y. and {Jean}, P. and {B{\oe}hm}, C. and {Kn{\"o}dlseder}, J.},
        title = "{Constraints on dark matter and the shape of the Milky Way dark halo from the 511-keV line}",
      journal = {\mnras},
         year = 2006,
        month = jun,
       volume = {368},
       number = {4},
        pages = {1695-1705},
          doi = {10.1111/j.1365-2966.2006.10226.x},
archivePrefix = {arXiv},
       eprint = {astro-ph/0507142},
 primaryClass = {astro-ph},
       adsurl = {https://ui.adsabs.harvard.edu/abs/2006MNRAS.368.1695A}
}

@ARTICLE{2015ApJ...807..130V,
       author = {{Venter}, C. and {Kopp}, A. and {Harding}, A.~K. and {Gonthier}, P.~L. and {B{\"u}sching}, I.},
        title = "{Cosmic-ray Positrons from Millisecond Pulsars}",
      journal = {\apj},
         year = 2015,
        month = jul,
       volume = {807},
       number = {2},
          eid = {130},
        pages = {130},
          doi = {10.1088/0004-637X/807/2/130},
archivePrefix = {arXiv},
       eprint = {1506.01211},
 primaryClass = {astro-ph.HE},
       adsurl = {https://ui.adsabs.harvard.edu/abs/2015ApJ...807..130V}
}
\bibliographystyle{spiejour}   % makes bibtex use spiejour.bst

%%%%% Biographies of authors %%%%%

% \vspace{1ex}
% \noindent Biographies and photographs of the other authors are not available.

% \listoffigures
% \listoftables

\end{spacing}
\end{document}